\documentclass[aps,amsmath,amssymb, reprint, prl, preprintnumbers]{revtex4-1}
\pdfoutput=1

\usepackage{graphicx}
\usepackage{epstopdf}
\usepackage{amsmath, amssymb}

\usepackage{hyperref}
\usepackage{bbm,array,amsfonts,graphicx,wrapfig,float,mathtools,multirow}
\usepackage[dvipsnames]{xcolor}

\newcommand{\be}{\begin{equation}}
\newcommand{\ee}{\end{equation}}
\newcommand{\beq}{\begin{equation}}
\newcommand{\beql}[1]{\begin{equation}\label{#1}}
\newcommand{\eeq}{\end{equation}}
\newcommand{\ba}{\begin{array}}
\newcommand{\ea}{\end{array}}
\newcommand{\bea}{\begin{eqnarray}}
\newcommand{\beal}[1]{\begin{eqnarray}\label{#1}}
\newcommand{\eea}{\end{eqnarray}}
\newcommand{\ben}{\begin{enumerate}}
\newcommand{\een}{\end{enumerate}}
\newcommand{\bean}{\begin{eqnarray*}}
\newcommand{\eean}{\end{eqnarray*}}
\newcommand{\eref}[1]{(\ref{#1})}

\newcommand{\nn}{\nonumber}

\newcommand{\fref}[1]{Figure \ref{#1}}

\begin{document}

\title{Birational Transformations on Dimer Integrable Systems}

\preprint{UNIST-MTH-25-RS-03, CGP25002}

\author{Minsung Kho}
\email[\texttt{minsung@unist.ac.kr}]{}
\affiliation{
Department of Mathematical Sciences, Ulsan National Institute of Science and Technology,
50 UNIST-gil, Ulsan 44919, South Korea
}

\author{Norton Lee}
\email[\texttt{norton.lee@ibs.re.kr}]{}
\affiliation{
Center for Geometry and Physics, Institute for Basic Science (IBS),
Pohang 37673, South Korea
}

\author{Rak-Kyeong Seong}
\email[\texttt{seong@unist.ac.kr}]{}
\affiliation{
Department of Mathematical Sciences, and Department of Physics, Ulsan National Institute of Science and Technology,
50 UNIST-gil, Ulsan 44919, South Korea
}

\begin{abstract}
We show that when two toric Calabi-Yau 3-folds 
and their corresponding toric varieties are related by a birational transformation, 
they are associated with a pair of dimer models on the 2-torus
that define dimer integrable systems, which themselves become birationally equivalent. 
Under this birational equivalence, 
the Casimirs and the Hamiltonias as well as the spectral curve and the Poisson structure formed by the cluster variables
of the dimer integrable systems are identified to each other. 
These integrable systems defined by dimer models were first introduced by Goncharov and Kenyon.
We illustrate this equivalence explicitly using a pair of dimer integrable systems corresponding to the abelian orbifolds of the form 
$\mathbb{C}^3/\mathbb{Z}_4 \times \mathbb{Z}_2$ with orbifold action $(1,0,3)(0,1,1)$ and 
$\mathcal{C}/\mathbb{Z}_2 \times \mathbb{Z}_2$ with action $(1,0,0,1)(0,1,1,0)$, 
whose spectral curves and Hamiltonians
are shown to be related by a birational transformation. 
\end{abstract} 

\maketitle


Birational geometry focuses on classifying algebraic varieties up to birational equivalence
and has profoundly impacted our understanding of algebraic varieties, particularly through the Minimal Model Program (MMP) initiated by Mori \cite{mori1, mori2, mori3}
and through subsequent developments \cite{Kawamata,zbMATH00232903, Kollar_Mori_1998, birkar2010existence, akhtar2012minkowski, gross2013birational, coates2016quantum, kasprzyk2017minimality, coates2021maximally, Coates2022MirrorSL}. 
More recently, it has been argued that such birational transformations
also connect supersymmetric gauge theories realized in string theory \cite{Ghim:2024asj, Ghim:2025zhs}.
In particular,  
when a D3-brane probes a toric Calabi-Yau 3-fold, its worldvolume theory is a $4d$ $\mathcal{N}=1$ supersymmetric quiver gauge theory,
which is encoded in a bipartite graph on a 2-torus also known as a dimer model \cite{KENYON1997591, kenyon2003introductiondimermodel} or a brane tiling \cite{Hanany:2005ve, Franco:2005rj, Franco:2005sm}.
It can be shown that if two such $4d$ $\mathcal{N}=1$ theories are related by a mass deformation \cite{Klebanov:1998hh, Gubser:1998ia, Bianchi:2014qma, Franco:2023tyf},
then the toric varieties of the associated toric Calabi-Yau 3-folds are related by a birational transformation \cite{akhtar2012minkowski, higashitani2022deformations, Ghim:2024asj, Ghim:2025zhs}. 

These birational transformations act on Newton polynomials of the form, 
\beal{es01a00}
P(x,y) = \sum_{(n_x,n_y) \in \Delta} c_{(n_x,n_y)} x^{n_x} y^{n_y}
\eea
where $x, y \in \mathbb{C}^*$
and $(n_x,n_y) \in \mathbb{Z}^2$ are coordinates of points in the toric diagram $\Delta$, which characterizes the toric variety $X(\Delta)$ \cite{fulton1993introduction, Leung:1997tw}. 
The Newton polynomial is also given by the permanent of the Kasteleyn matrix $K$ of the dimer model, 
$P(x,y) = \text{perm}(K)$ \cite{kenyon2003introductiondimermodel, kasteleyn1967graph}.
The birational transformations $\varphi_A$ act on $P(x,y)$ as follows, 
\beal{es01a01}
\varphi_A ~:~
(x,y) &\mapsto& (A(y) x, y)
\eea
where the Laurent polynomial $A(y)$ is chosen to be such that 
$A(y)^{-m}$ is a polynomial divisor of $C_{m}(y)$ for $a \leq m \leq -1$ in the expansion
$P(x,y) = \sum_{m=a}^{b} C_m(y) x^m$.
Here, $a<0$ and $b>0$ and $C_m (y)$ are sub-polynomials for $a \leq m \leq b$.
We identify the corresponding toric varieties as birationally equivalent, if the birational map in \eref{es01a01} applies
for at least one chosen $GL(2,\mathbb{Z})$ frame or choice of origin in $\mathbb{Z}^2$ for the toric diagrams $\Delta$.

As shown by Goncharov and Kenyon \cite{goncharov2012dimersclusterintegrablesystems, EagerFrancoSchaeffer}, 
a dimer model realized as a bipartite graph on $T^2$ gives rise to an integrable system \cite{EagerFrancoSchaeffer, Bershtein:2017swf, Marshakov:2019vnz, Huang:2020neq, Lee:2023wbf, Lee:2024bqg}, 
where the spectral curve is given by the zero locus, 
\beal{es01a05}
\Sigma ~:~ P(x,y) = 0 ~.~
\eea
This holomorphic curve also plays an essential role in the Type IIB brane configuration consisting of a D5-brane suspended between an NS5-brane wrapping $\Sigma$,
which is represented by the bipartite graph of the dimer model and is T-dual to the D3-brane probing the toric Calabi-Yau 3-fold \cite{Douglas:1997de, Douglas:1996sw, Feng:2000mi, Feng:2001xr}. 
$\Sigma$ is also known as the mirror curve 
of the mirror Calabi-Yau 3-fold \cite{Hori:2000ck, Hori:2000kt, Feng:2005gw}. 

In this work, we discover for the first time that the integrable systems defined by two dimer models 
whose toric Calabi-Yau 3-folds are related by a birational transformation as given in \eref{es01a01}, 
are identical up to a redefinition of parameters corresponding to zig-zag paths and face loops in the dimer models. 
We illustrate this discovery with an explicit example
based on the dimer models corresponding to the
abelian orbifold of the form $\mathbb{C}^3/\mathbb{Z}_4 \times \mathbb{Z}_2$ with orbifold action $(1,0,3)(0,1,1)$ \cite{Davey:2010px, Hanany:2010ne}, also referred to as Model 2 in the classification of \cite{Hanany:2012hi},
and toric phase a of the abelian orbifold of the form $\mathcal{C}/\mathbb{Z}_2 \times \mathbb{Z}_2$ with action $(1,0,0,1)(0,1,1,0)$, also referred to as phase a of $\text{PdP}_{5}$ \cite{Feng:2002fv,Franco:2005rj} and Model 4a in \cite{Hanany:2012hi}.

\section{Dimer Integrable Systems}

A perfect matching $p_a$ is a set of edges $e_{ij} = (w_i, b_j)$ in the dimer model, 
such that edges in $p_a$ connect to all white nodes $w_i$ and black nodes $b_i$ uniquely once. 
These perfect matchings play an important role as GLSM fields \cite{Witten:1993yc} that parameterize the associated toric Calabi-Yau 3-fold. 
We define the weight $\overline{p}_a$ of a perfect matching $p_a$
as the product,
\beal{es03a00}
\overline{p}_a = \prod_{e_{ij} \in p_a} e_{ij}^{+}
~,~
(\overline{p}_a)^{-1} = \prod_{e_{ij} \in p_a} e_{ij}^{-}
~,~
\eea
where we introduce directed edges defined as $e_{ij}^{+} ~:~ w_i \rightarrow b_j$ 
and
$e_{ij}^{-} ~:~ b_j \rightarrow w_i$.
Here, the positive sign indicates an edge directed from a white node $w_i$ to a black node $b_j$,
and the negative sign indicates the same edge but in the opposite direction.
With this convention, we define the following product of perfect matching weights,
\beal{es03a000}
\overline{p}_a \cdot (\overline{p}_b)^{-1}
\equiv 
 \cdots~e_{ij}^{+}~e_{kj}^{-}~e_{kl}^{+}~e_{ml}^{-}~\cdots ~,~
\eea
where $\overline{p}_a= \cdots ~ e_{ij}^{+} ~e_{kl}^{+}~\cdots $ and $(\overline{p}_b)^{-1}= \cdots ~ e_{kj}^{-} ~e_{ml}^{-}~\cdots $.
This product yields an ordered sign-alternating sequence of directed edge variables corresponding to a path in the dimer. 

All closed connected and directed paths composed of sign-alternating edges $e_{ij}^\pm$ in a dimer model on $T^2$ can be expressed as permutation tuples of the permutation group $S_{2n_e}$, where $n_e$ is the number of edges in the dimer model
\cite{Jejjala:2010vb, Hanany:2015tgh}.
Given two such permutation tuples, we can define the following product,
\beal{es03a02}
&&
( \cdots e_{ik}^{-}~ e_{ij}^{+} ~e_{mj}^{-} \cdots)  ( \cdots e_{uj}^{+}~ e_{ij}^{-} ~e_{iv}^{+} \cdots)
\nn\\
&&
=
( \cdots e_{ik}^{-}~e_{iv}^{+} \cdots) ( \cdots e_{uj}^{+}~e_{mj}^{-} \cdots)   
~,~
\eea
giving a new pair of closed paths
with the identities
$(e_{ij}^\pm)^{-1} = e_{ij}^\mp$ and
$e_{ij}^{+} ~e_{ij}^{-} = 1$.

\subsection{Model 2: $\mathbb{C}^3/\mathbb{Z}_4 \times \mathbb{Z}_2$ $(1,0,3)(0,1,1)$}

\begin{figure}[ht!!]
\begin{center}
\resizebox{0.97\hsize}{!}{
\includegraphics[height=5cm]{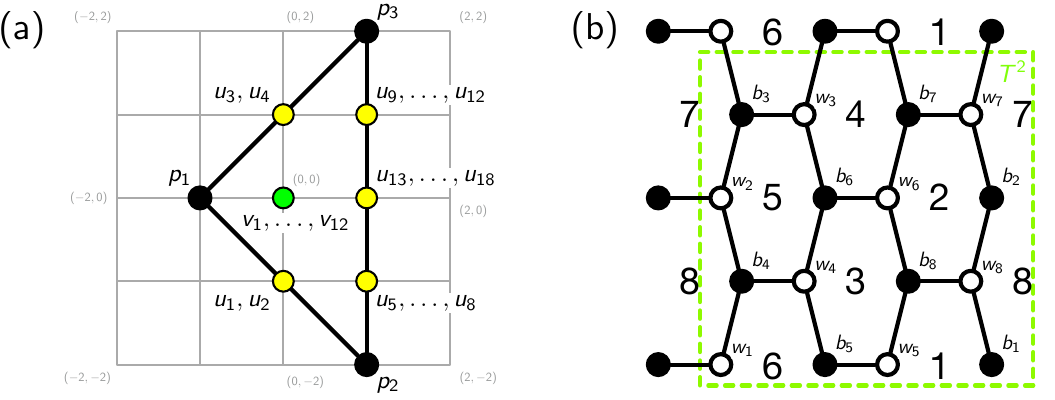} 
}
\caption{
(a) The toric diagram and (b) the dimer model for $\mathbb{C}^3/\mathbb{Z}_4 \times \mathbb{Z}_2$ $(1,0,3)(0,1,1)$ [Model 2]. 
Points in the toric diagram are labelled by corresponding perfect matchings in the dimer model.
\label{f_fig01}}
 \end{center}
 \end{figure}

The dimer model, referred to here as Model 2, corresponding to the abelian orbifold of the form $\mathbb{C}^3/\mathbb{Z}_4 \times \mathbb{Z}_2$ with orbifold action $(1,0,3)(0,1,1)$
is shown in \fref{f_fig01}(b) with a choice of a fundamental domain on $T^2$.
The corresponding Kasteleyn matrix takes the following form, 
\beal{es04a00}
K
=
\resizebox{0.35\textwidth}{!}{$
\left(
\ba{c|cccccccc}
\; & b_1 & b_2 & b_3 & b_4 & b_5 & b_6 & b_7 & b_8
\\
\hline
w_1    &   e_{11} x^{-1} & 0 & e_{13} y^{-1} & e_{14} & 0 & 0 & 0 & 0 
\\
w_2.   &  0 & e_{22} x^{-1} & e_{23} & e_{24} & 0 & 0 & 0 & 0 
\\
w_3    &  0 & 0 & e_{33} & 0 & e_{35} y & e_{36} & 0 & 0 
\\
w_4    &  0 & 0 & 0 & e_{44} & e_{45} & e_{46} & 0 & 0 
\\
w_5    &  0 & 0 & 0 & 0 & e_{55} & 0 & e_{57} y^{-1} & e_{58}
\\
w_6    &  0 & 0 & 0 & 0 & 0 & e_{66} & e_{67} & e_{68} 
\\
w_7    &  e_{71} y & e_{72} & 0 & 0 & 0 & 0 & e_{77} & 0 
\\
w_{8} &  e_{81} & e_{82} & 0 & 0 & 0 & 0 & 0 & e_{88} 
\\
\ea
\right)
$}
~.~
\eea
The associated toric diagram can be shifted in the $(1,0)$
direction as illustrated in \fref{f_fig01}(a) such that the new corresponding Newton polynomial takes the form, 
\beal{es04a00b2}
&&
P(x,y)
=
\overline{p}_1 
\cdot \Big[
\delta_{(-1,0)} \frac{1}{x} 
+ \delta_{(1,-2)} \frac{x}{y^2}
+ \delta_{(1,2)} x y^2
\nn\\
&& 
\hspace{0.5cm}
+ \delta_{(0,-1)} \frac{1}{y}
+ \delta_{(0,1)} y
+ \delta_{(1,-1)} \frac{x}{y}
+ \delta_{(1,1)} x y
\nn\\
&& 
\hspace{0.5cm}
+ \delta_{(1,0)} x
+ H
\Big]~,~
\eea
where the reference perfect matching weight $\overline{p}_1=e_{11}^{+} e_{22}^{+} e_{33}^{+} e_{44}^{+} e_{55}^{+} e_{66}^{+} e_{77}^{+} e_{88}^{+}$
has been factored out.
This allows us to identify
the coefficients with Casimir loops $\delta_{(n_x,n_y)}$, which correspond to points $(n_x,n_y)$ in the toric diagram in \fref{f_fig01}(a). 
Additionally, we obtain the constant term $H$, which corresponds to the Hamiltonian of the dimer integrable system that can be expressed as a sum over all 1-loops, $H=\sum_{u=1}^{12} \gamma_u$.

The 1-loops $\gamma_{1, \dots, 12}$ and the Casimir loops $\delta_{(n_x,n_y)}$
can be expressed in terms of
zig-zag paths $z_{1,\dots, 8}$,
\beal{es04a01}
&
z_1 = (e_{14}^{+}~e_{44}^{-}~e_{46}^{+}~e_{66}^{-}~e_{67}^{+}~e_{77}^{-}~e_{71}^{+}~e_{11}^{-})~,~
&
\nn\\
&
z_2 = (e_{23}^{+}~e_{33}^{-}~e_{35}^{+}~e_{55}^{-}~e_{58}^{+}~e_{88}^{-}~e_{82}^{+}~e_{22}^{-})~,~
&
\nn\\
&
z_3 = (e_{11}^{+}~e_{81}^{-}~e_{88}^{+}~e_{68}^{-}~e_{66}^{+}~e_{36}^{-}~e_{33}^{+}~e_{13}^{-} )~,~
&
\nn\\
&
z_4 = (e_{22}^{+}~e_{72}^{-}~e_{77}^{+}~e_{57}^{-}~e_{55}^{+}~e_{45}^{-}~e_{44}^{+}~e_{24}^{-} )~,~
&
\nn\\
&
z_5 = (e_{13}^{+}~e_{23}^{-}~e_{24}^{+}~e_{14}^{-})~,~
z_6 = (e_{36}^{+}~e_{46}^{-}~e_{45}^{+}~e_{35}^{-})~,~
&
\nn\\
&
z_7 = (e_{57}^{+}~e_{67}^{-}~e_{68}^{+}~e_{58}^{-})~,~
z_8 = (e_{72}^{+}~e_{82}^{-}~e_{81}^{+}~e_{71}^{-})~,~
&
\eea
and the face loops $f_{1,\dots,8}$
\beal{es04a02}
f_1 &=& (e_{57}^{+}~e_{77}^{-}~e_{71}^{+}~e_{81}^{-}~e_{88}^{+}~e_{58}^{-})~,~\nn\\
f_2 &=& (e_{68}^{+}~e_{88}^{-}~e_{82}^{+}~e_{72}^{-}~e_{77}^{+}~e_{67}^{-})~,~\nn\\
f_3 &=& (e_{45}^{+}~e_{55}^{-}~e_{58}^{+}~e_{68}^{-}~e_{66}^{+}~e_{46}^{-})~,~\nn\\
f_4 &=& (e_{36}^{+}~e_{66}^{-}~e_{67}^{+}~e_{57}^{-}~e_{55}^{+}~e_{35}^{-})~,~\nn\\
f_5 &=& (e_{24}^{+}~e_{44}^{-}~e_{46}^{+}~e_{36}^{-}~e_{33}^{+}~e_{23}^{-})~,~\nn\\
f_6 &=& (e_{13}^{+}~e_{33}^{-}~e_{35}^{+}~e_{45}^{-}~e_{44}^{+}~e_{14}^{-})~,~\nn\\
f_7 &=& (e_{11}^{+}~e_{71}^{-}~e_{72}^{+}~e_{22}^{-}~e_{23}^{+}~e_{13}^{-})~,~\nn\\
f_8 &=& (e_{22}^{+}~e_{82}^{-}~e_{81}^{+}~e_{11}^{-}~e_{14}^{+}~e_{24}^{-})~,~
\eea
of the dimer model.
The face loops are the cluster variables of the quiver associated to the dimer model. 
The zig-zag paths and face loops form the following relations, 
\beal{es04a03}
&
f_1 f_2 = z_7 z_8^{-1} ~,~
f_3 f_4 = z_6 z_7^{-1} ~,~
&
\nn\\
&
f_5 f_6  = z_5 z_6^{-1} ~,~
f_7 f_8 = z_5^{-1} z_8 ~,~
&
\nn\\
&
f_1 f_3 f_5  f_7 = z_3 z_4^{-1}~,~
f_2 f_4 f_6  f_8 = z_4 z_3^{-1}~,~
&
\nn\\
&
f_2 f_3 f_6 f_7 = z_1^{-1} z_2~,~
f_1 f_4 f_5 f_8 = z_1 z_2^{-1}~.~
&
\eea
We note here that there are 2 degrees of freedom corresponding to the pair of canonical variables $e^{P}$ and $e^{Q}$, 
which allows us to write the face loops as follows,
\beal{es04a04}
f_1 &=& 
e^{P}
z_1^{1/2} z_2^{-1/2} z_7^{1/2} z_8^{-1/2}
~,~
\nn\\
f_2 &=& 
e^{-P}
z_1^{-1/2} z_2^{1/2} z_7^{1/2} z_8^{-1/2}
~,~
\nn\\
f_3 &=& 
e^{Q}
z_1^{-1/2} z_2^{1/2} z_6^{1/2} z_7^{-1/2}
~,~
\nn\\
f_4 &=& 
e^{-Q}
z_1^{1/2} z_2^{-1/2} z_6^{1/2} z_7^{-1/2}
~,~
\nn\\
f_5 &=& 
e^{-P}
z_3^{1/2} z_4^{-1/2} z_5^{1/2} z_6^{-1/2}
~,~
\nn\\
f_6 &=& 
e^{P}
z_3^{-1/2} z_4^{1/2} z_5^{1/2} z_6^{-1/2}
~,~
\nn\\
f_7 &=& 
e^{-Q}
z_3^{1/2} z_4^{-1/2} z_5^{-1/2} z_8^{1/2}
~,~
\nn\\
f_8 &=& 
e^{Q}
z_3^{-1/2} z_4^{1/2} z_5^{-1/2} z_8^{1/2}
~,~
\eea
where $z_1 z_2 z_3 z_4 z_5 z_6 z_7 z_8 = 1$.

In terms of these zig-zag paths and face loops, 
the 1-loops $\gamma_{1, \dots, 12}$ can be expressed as,
\beal{es04a05}
&
\gamma_1 = z_1 z_8 f_2 ~,~
\gamma_2 = z_1 z_8 f_2 f_3~,~
\gamma_3 = z_1 z_8  f_1 f_2 f_3~,~
&
\nn\\
&
\gamma_4 = z_1 z_8 f_2 f_3 f_6~,~
\gamma_5 = z_1 z_8 f_1 f_2 f_3 f_6~,~
&
\nn\\
&
\gamma_6 = z_1 z_8 f_1 f_2 f_3 f_4 f_6~,~
\gamma_7 = z_1 z_8 f_1 f_2 f_3 f_6 f_7~,~
&
\nn\\
&
\gamma_8 = z_1 z_8 f_1 f_2 f_3 f_4 f_6 f_7~,~
\gamma_9 = z_1 z_8 f_1 f_2^2 f_3 f_4 f_6 f_7~,~
&
\nn\\
&
\gamma_{10} = z_1 z_8 f_1 f_2 f_3 f_4 f_5 f_6 f_7~,~
&
\nn\\
&
\gamma_{11} = z_1 z_8 f_1 f_2^2 f_3 f_4 f_5 f_6 f_7~,~
&
\nn\\
&
\gamma_{12} = z_1 z_8 f_1 f_2^2 f_3^2 f_4 f_5 f_6 f_7~,~
&
\eea
and the Casimir loops $\delta_{(n_x,n_y)}$ can be expressed as,
\beal{es04a10}
&
\delta_{(1,-2)} = z_3^{-1} z_4^{-1} ~,~
\delta_{(1,2)} = z_1 z_2 ~,~
\delta_{(-1,0)} = 1 ~,~
&
\nn\\
&
\delta_{(0,-1)}^{1} = z_3^{-1} ~,~
\delta_{(0,-1)}^{2} = z_4^{-1} ~,~
&
\nn\\
&
\delta_{(0,1)}^{1} = z_1 ~,~
\delta_{(0,1)}^{2} = z_2 ~,~
&
\nn\\
&
\delta_{(1,-1)}^{1} = z_3^{-1} z_4^{-1} z_5^{-1} ~,~
\delta_{(1,-1)}^{2} = z_3^{-1} z_4^{-1} z_6^{-1} ~,~
&
\nn\\
&
\delta_{(1,-1)}^{3} = z_3^{-1} z_4^{-1} z_7^{-1} ~,~
\delta_{(1,-1)}^{4} = z_3^{-1} z_4^{-1} z_8^{-1} ~,~
&
\nn\\
&
\delta_{(1,1)}^{1} = z_1 z_2 z_5 ~,~
\delta_{(1,1)}^{2} = z_1 z_2 z_6 ~,~
&
\nn\\
&
\delta_{(1,1)}^{3} = z_1 z_2 z_7 ~,~
\delta_{(1,1)}^{4} = z_1 z_2 z_8 ~,~
&
\nn\\
&
\delta_{(1,0)}^{1} = z_1 z_2 z_5 z_6 ~,~
\delta_{(1,0)}^{2} = z_1 z_2 z_6 z_7 ~,~
&
\nn\\
&
\delta_{(1,0)}^{3} = z_1 z_2 z_5 z_7 ~,~
\delta_{(1,0)}^{4} = z_1 z_2 z_5 z_8 ~,~
&
\nn\\
&
\delta_{(1,0)}^{5} = z_1 z_2 z_6 z_8 ~,~
\delta_{(1,0)}^{6} = z_1 z_2 z_7 z_8 ~.~
&
\eea
Like perfect matchings, multiple Casimir loops can correspond to the same point $(n_x,n_y)$ in the toric diagram in \fref{f_fig01}(a)
and therefore can be combined as follows,
\beal{es04a15}
\delta_{(0,-1)} 
&=& 
\delta_{(0,-1)}^{1} + \delta_{(0,-1)}^{2}
= z_3^{-1} + z_4^{-1} 
~,~ \nn\\
\delta_{(1,-1)} 
&=&
\delta_{(1,-1)} ^{1} + \delta_{(1,-1)} ^{2} + \delta_{(1,-1)} ^{3} + \delta_{(1,-1)} ^{4} 
\nn\\
&=&
z_3^{-1} z_4^{-1} (z_5^{-1} + z_6^{-1} + z_7^{-1} + z_8^{-1})
~,~
\nn\\
\delta_{(1,0)}
&=&
\delta_{(1,0)}^{1} + \delta_{(1,0)}^{2} + \delta_{(1,0)}^{3}
\nn\\
&&
+\delta_{(1,0)}^{4} + \delta_{(1,0)}^{5} + \delta_{(1,0)}^{6}
~,~
\nn\\
&=&
z_1 z_2 (z_5 z_6 + z_5 z_7 + z_5 z_8 
\nn\\
&&
\hspace{1cm}
+ z_6 z_7 + z_6 z_8 + z_7 z_8)
~,~
\nn\\
\delta_{(1,1)}
&=&
\delta_{(1,1)}^{1} + \delta_{(1,1)}^{2} + \delta_{(1,1)}^{3} + \delta_{(1,1)}^{4}
\nn\\
&=&
z_1 z_2 (z_5 + z_6 + z_7 + z_8)
~,~
\nn\\
\delta_{(0,1)}
&=&
\delta_{(0,1)}^{1} + \delta_{(0,1)}^{2}
= z_1 + z_2
~.~
\eea

The zero locus of the Newton polynomial in \eref{es04a00b2} gives the spectral curve of the dimer integrable system for Model 2, 
\beal{es04a16}
&&
\Sigma ~:~
(\frac{z_5}{y}  + 1)(\frac{z_6}{y}  + 1) (\frac{z_7}{y}  + 1) (\frac{z_8}{y}  + 1) z_1 z_2 x y^2
\nn\\
&&
\hspace{0.5cm}
+ \Big(\frac{1}{z_3} + \frac{1}{z_4}\Big) \frac{1}{y} + (z_1 + z_2) y
+ \frac{1}{x} + H = 0
~.~
\eea

\subsection{Model 4a: $\mathcal{C}/\mathbb{Z}_2 \times \mathbb{Z}_2$ $(1,0,0,1) (0,1,1,0)$}

\begin{figure}[ht!!]
\begin{center}
\resizebox{0.97\hsize}{!}{
\includegraphics[height=5cm]{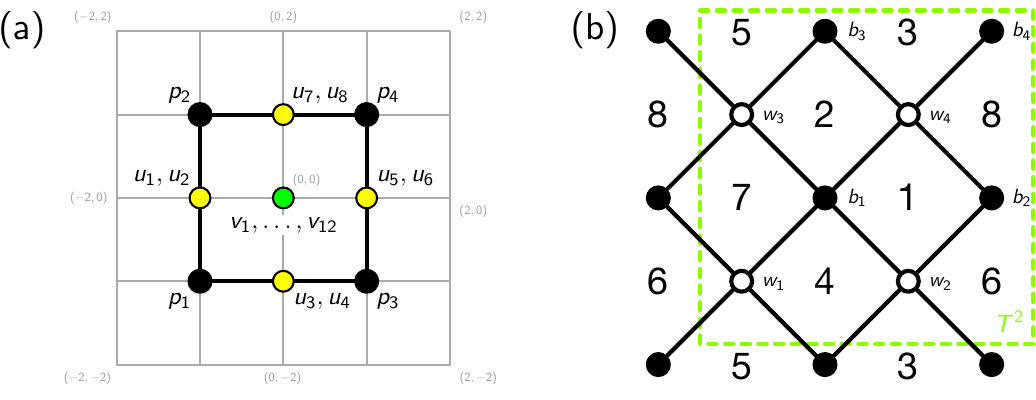} 
}
\caption{
(a) The toric diagram and (b) the dimer model for $\mathcal{C}/\mathbb{Z}_2 \times \mathbb{Z}_2$ $(1,0,0,1) (0,1,1,0)$ [Model 4a]. 
Points in the toric diagram are labelled by corresponding perfect matchings in the dimer model.
\label{f_fig02}}
 \end{center}
 \end{figure}

The dimer model corresponding to the abelian orbifold of the form $\mathcal{C}/\mathbb{Z}_2 \times \mathbb{Z}_2$ with orbifold action $(1,0,0,1) (0,1,1,0)$
is shown in \fref{f_fig02}(b).
The corresponding Kasteleyn matrix is given by,
\beal{es06a00}
K=
\resizebox{0.24\textwidth}{!}{$
\left(
\ba{c|cccc}
\; & b_1 & b_2 & b_3 & b_4
\\
\hline
w_1 & e_{11} & e_{12} x^{-1} & e_{13} y^{-1} & e_{14} x^{-1} y^{-1}
\\
w_2 & e_{21}  & e_{22} & e_{23} y^{-1} & e_{24} y^{-1}
\\
w_3 & e_{31}  & e_{32} x^{-1} & e_{33} & e_{34} x^{-1}
\\
w_4 & e_{41}  & e_{42}  & e_{43}  & e_{44}
\\
\ea
\right)
$}
~.~
\eea
When we shift the toric diagram in the $(1,1)$ direction resulting in the toric diagram in \fref{f_fig02}(a), the corresponding Newton polynomial becomes,
\beal{es06a00b2}
&&
P(x,y)
= \overline{u}_1 \cdot
\Big[
\delta_{(-1,-1)} \frac{1}{x y} 
+ \delta_{(-1,1)} \frac{y}{x} 
+ \delta_{(1,-1)} \frac{x}{y} 
\nn\\
&& \hspace{0.5cm}
+ \delta_{(1,1)} xy
+ \delta_{(-1,0)} \frac{1}{x} 
+ \delta_{(0,-1)} \frac{1}{y} 
+ \delta_{(1,0)} x 
\nn\\
&& \hspace{0.5cm}
+ \delta_{(0,1)} y
+ H
\Big]
~,~
\eea
where the reference perfect matching weight $\overline{u}_1=e_{14}^{+} e_{21}^{+} e_{32}^{+} e_{43}^{+}$
has been factored out in order for the coefficients to become Casimir loops $\delta_{(n_x,n_y)}$.
These correspond to points in the toric diagram in \fref{f_fig02}(a). 
We also obtain the Hamiltonian $H$, which can be expressed as a sum over all 1-loops, $H=\sum_{u=1}^{12} \gamma_u$.

The 1-loops $\gamma_{1, \dots, 12}$ and the Casimir loops $\delta_{(n_x,n_y)}$
can be expressed in terms of zig-zag paths $z_{1, \dots, 8}$,
\beal{es06a01}
&
z_1 = (e_{11}^{+}~e_{21}^{-}~e_{22}^{+}~e_{12}^{-})~,~
z_2 = (e_{32}^{+}~e_{42}^{-} ~e_{41}^{+}~e_{31}^{-})~,~
&
\nn\\
&
z_3 = (e_{33}^{+}~e_{43}^{-}~e_{44}^{+}~e_{34}^{-})~,~
z_4 = (e_{14}^{+}~e_{24}^{-} ~e_{23}^{+}~e_{13}^{-})~,~
&
\nn\\
&
z_5 = (e_{13}^{+}~e_{33}^{-}~e_{31}^{+}~e_{11}^{-})~,~
z_6 = (e_{21}^{+}~e_{41}^{-}~e_{43}^{+}~e_{23}^{-})~,~
&
\nn\\
&
z_7 = (e_{24}^{+}~e_{44}^{-}~e_{42}^{+}~e_{22}^{-})~,~
z_8 = (e_{12}^{+}~e_{32}^{-}~e_{34}^{+}~e_{14}^{-})~,~
&
\eea
and the face loops $f_{1,\dots,8}$,
\beal{es04a02}
&
f_1 = (e_{22}^{+}~e_{42}^{-}~e_{41}^{+}~e_{21}^{-})~,~
f_2 = (e_{43}^{+}~e_{33}^{-}~e_{31}^{+}~e_{41}^{-})~,~
&
\nn\\
&
f_3 = (e_{44}^{+}~e_{24}^{-}~e_{23}^{+}~e_{43}^{-})~,~
f_4 = (e_{21}^{+}~e_{11}^{-}~e_{13}^{+}~e_{23}^{-})~,~
&
\nn\\
&
f_5 = (e_{33}^{+}~e_{13}^{-}~e_{14}^{+}~e_{34}^{-})~,~
f_6 = (e_{12}^{+}~e_{22}^{-}~e_{24}^{+}~e_{14}^{-})~,~
&
\nn\\
&
f_7 = (e_{11}^{+}~e_{31}^{-}~e_{32}^{+}~e_{12}^{-})~,~
f_8 = (e_{34}^{+}~e_{44}^{-}~e_{42}^{+}~e_{32}^{-})~.~
&
\eea
The face loops and zig-zag paths form the following relations, 
\beal{es06a03}
&
f_1 f_7 = z_1 z_2 ~,~
f_2 f_8 = z_2^{-1} z_3^{-1} ~,~
&
\nn\\
&
f_3 f_5 = z_3 z_4 ~,~
f_4 f_6 = z_1^{-1} z_4^{-1} ~,~
&
\nn\\
&
f_2 f_4 = z_5 z_6 ~,~
f_1 f_3 = z_6^{-1} z_7^{-1} ~,~
&
\nn\\
&
f_6 f_8 = z_7 z_8 ~,~
f_5 f_7 = z_5^{-1} z_{8}^{-1} ~.~
&
\eea
Here, we note that there are 2 degrees of freedom corresponding to the pair of canonical variables $e^P$ and $e^Q$, 
which allows us to express the face loops as follows,
\beal{es06a04}
&
f_1 =
e^{Q} ~,~
f_2 =
e^{P} ~,~
&
\nn\\
&
f_3 =
e^{-Q} z_6^{-1} z_7^{-1} 
~,~
f_4 =
e^{-P} z_5 z_6 
~,~
&
\nn\\
&
f_5 =
e^{Q} z_3 z_4 z_6 z_7
~,~
f_6 =
e^{P} z_1^{-1} z_4^{-1} z_5^{-1} z_6^{-1}
~,~
&
\nn\\
&
f_7 =
e^{-Q} z_1 z_2
~,~
f_8 = 
e^{-P} z_2^{-1} z_3^{-1}
~,~
&
\eea
where $z_1 z_2 z_3 z_4 z_5 z_6 z_7 z_8=1$.

In terms of the face loops and the zig-zag paths, we can express the 1-loops $\gamma_{1, \dots, 12}$ as,
\beal{es06a05}
&
\gamma_1 = z_1 f_6 ~,~
\gamma_2 = z_1 f_3 f_6 ~,~
\gamma_3 = z_1 f_3 f_6 f_8 ~,~
&
\nn\\
&
\gamma_4 = z_1 f_3 f_4 f_6 ~,~
\gamma_5 = z_1 f_3 f_4 f_6 f_8 ~,~
&
\nn\\
&
\gamma_6 = z_1 f_1 f_3 f_4 f_6 f_8 ~,~
\gamma_7 = z_1 f_3 f_4 f_5 f_6 f_8 ~,~
&
\nn\\
&
\gamma_8 = z_1 f_1 f_3 f_4 f_5 f_6 f_8 ~,~
\gamma_9 = z_1 f_1 f_3 f_4 f_5 f_6^2 f_8 ~,~
&
\nn\\
&
\gamma_{10} = z_1 f_1 f_2 f_3 f_4 f_5 f_6 f_8 ~,~
\gamma_{11} = z_1 f_1 f_2 f_3 f_4 f_5 f_6^2 f_8 ~,~
&
\nn\\
&
\gamma_{12} = z_1 f_1 f_2 f_3^2 f_4 f_5 f_6^2 f_8 ~,~
&
\eea
and the Casimir loops $\delta^a_{(n_x,n_y)}$ as,
\beal{es06a10}
&
\delta_{(-1,-1)} = z_6^{-1} ~,~
\delta_{(-1,1)} = z_8 ~,~
&
\nn\\
&
\delta_{(1,-1)} = z_2^{-1} z_4^{-1} z_6^{-1} ~,~
\delta_{(1,1)} = z_1 z_3 z_8 ~,~
&
\nn\\
&
\delta_{(-1,0)}^{1} = 1 ~,~
\delta_{(-1,0)}^{2} = z_6^{-1} z_8  ~,~
&
\nn\\
&
\delta_{(0,-1)}^{1} = z_2^{-1} z_6^{-1} ~,~
\delta_{(0,-1)}^{2} = z_4^{-1} z_6^{-1} ~,~
&
\nn\\
&
\delta_{(1,0)}^{1} = z_1 z_3 z_7 z_8 ~,~
\delta_{(1,0)}^{2} = z_2^{-1} z_4^{-1} z_6^{-1} z_7^{-1} ~,~
&
\nn\\
&
\delta_{(0,1)}^{1} = z_1 z_8 ~,~
\delta_{(0,1)}^{2} = z_3 z_8 ~,~
&
\eea
where Casimir loops corresponding to the same point in the toric diagram in \fref{f_fig02}(a) can be combined as follows, 
\beal{es06a15}
\delta_{(-1,0)} &=&
\delta_{(-1,0)}^{1} + \delta_{(-1,0)}^{2} 
= 1 + z_6^{-1} z_8
~,~
\nn\\
\delta_{(0,-1)} &=&
\delta_{(0,-1)}^{1} + \delta_{(0,-1)}^{2}
= z_2^{-1} z_6^{-1} + z_4^{-1} z_6^{-1}
~,~
\nn\\
\delta_{(1,0)} &=&
\delta_{(1,0)}^{1} + \delta_{(1,0)}^{2}
=  z_1 z_3 z_7 z_8 + z_2^{-1} z_4^{-1} z_6^{-1} z_7^{-1}
~,~
\nn\\
\delta_{(0,1)} &=&
\delta_{(0,1)}^{1} + \delta_{(0,1)}^{2}
= z_1 z_8 + z_3 z_8 ~.~
\eea

The spectral curve of the dimer integrable system for Model 4a
is given by the zero locus of the Newton polynomial in \eref{es06a00b2},
\beal{es06a16}
&&
\Sigma ~:~
\Big(\frac{1}{y} + z_6\Big) \Big(\frac{1}{y}+z_8\Big)  \frac{y}{z_6  x}
+\Big(\frac{1}{z_2} + \frac{1}{z_4}\Big) \frac{1}{z_6 y}
\nn\\
&&
\hspace{0.5cm}
+ \Big(\frac{1}{y}+ \frac{1}{z_5}\Big) \Big(\frac{1}{y}+\frac{1}{z_7}\Big) \frac{ x y }{z_2 z_4 z_6 } 
+(z_1 + z_3) z_8 y 
\nn\\
&&
\hspace{0.5cm}
+ H = 0 ~.~
\eea

\section{Birational Equivalence}

Let us refer to the Newton polynomial in \eref{es04a00b2}
and the spectral curve in \eref{es04a16} for Model 2 as
$P^{(2)}(x,y)$ and $\Sigma^{(2)}$, respectively, 
in order to distinguish them clearly from the Newton polynomial in \eref{es06a00b2}
and the spectral curve in \eref{es06a16} for Model 4a, which we denote by $P^{(4a)}(x,y)$ and $\Sigma^{(4a)}$, respectively.
The following birational transformation,
\beal{es10a10}
\varphi_A ~:~ 
(x,y) \mapsto 
\left(
\left(
1+ \frac{1}{y}
\right)^2
xy
~,~
y
\right)
~,~
\eea
acts on the Newton polynomial 
with all perfect matching weights set to 1, 
yielding the identification, 
\beal{es10a10b}
\varphi_A 
P^{(4a)}(x,y) = P^{(2)}(x,y) ~.~
\eea
This birational transformation relates 
the toric varieties
associated to Models 2 and 4a.

Under a refined version of the birational transformation in \eref{es10a10}, given by,
\beal{es10a11}
&&
\varphi_A ~:~ 
(x,y) \mapsto 
\nn\\
&&
\hspace{0.5cm}
\left(
\left(
z_6^{(4a)} + \frac{1}{y}
\right)
\left(
z_8^{(4a)} + \frac{1}{y}
\right)
\frac{y}{z_6^{(4a)}}
x
~,~
y
\right)
~,~
\eea
with $A(y)= (z_6^{(4a)} + \frac{1}{y})(z_8^{(4a)} + \frac{1}{y}) \frac{y}{z_6^{(4a)}}$,
we discover that the spectral curve $\Sigma^{(4a)}$ in \eref{es06a16} 
is mapped to $\Sigma^{(2)}$ in \eref{es04a16},
\beal{es10a12}
\varphi_A \Sigma^{(4a)} = \Sigma^{(2)} ~.~
\eea
Under this identification, 
the zig-zag loops
satisfy, 
\beal{es10a25}
&
z_1^{(4a)} z_8^{(4a)} = z_1^{(2)} ~,~
z_3^{(4a)} z_8^{(4a)} = z_2^{(2)} ~,~
z_5^{(4a)} = z_5^{(2)}  ~,~
&
\nn\\
&
z_{2}^{(4a)} z_6^{(4a)} = z_{4}^{(2)} ~,~
z_{4}^{(4a)} z_6^{(4a)} = z_{3}^{(2)} ~,~
z_6^{(4a)} = (z_6^{(2)})^{-1} ~,~
&
\nn\\
&
z_7^{(4a)} = z_7^{(2)}  ~,~
z_8^{(4a)} = (z_8^{(2)})^{-1} ~,~
&
\eea
and the face loops between the two dimer models are identified to each other as follows,
\beal{es10a30}
&
f_8^{(4a)} = f_1^{(2)} ~,~ 
f_6^{(4a)} = f_2^{(2)} ~,~ 
f_3^{(4a)} = f_3^{(2)} ~,~ 
&
\nn\\
&
f_1^{(4a)} = f_4^{(2)} ~,~
f_2^{(4a)} = f_5^{(2)} ~,~ 
f_4^{(4a)} = f_6^{(2)} ~,~ 
&
\nn\\
&
f_5^{(4a)} = f_7^{(2)} ~,~ 
f_7^{(4a)} = f_8^{(2)} ~.~
&
\eea

Moreover, the 1-loops of the two integrable systems also obey,
\beal{es10a31}
\gamma_u^{(4a)} = \gamma_u^{(2)} ~,~
\eea
for all $u=1, \dots, 12$. 
This implies that the two Hamiltonians of the dimer integrable systems are identical under the birational transformation in \eref{es10a11},
\beal{es10a31}
H^{(4a)} = H^{(2)} ~.~
\eea
By identifying \eref{es04a04} with \eref{es06a04}, we also obtain the following canonical transformation, 
\beal{es10a32}
e^{Q^{(4a)}} &=& e^{-Q^{(2)}} (z_1^{(2)}z_6^{(2)})^{1/2}(z_2^{(2)}z_7^{(2)})^{-1/2}
~,~
\nn\\
e^{P^{(4a)}} &=& e^{-P^{(2)}} (z_3^{(2)}z_5^{(2)})^{1/2}(z_4^{(2)}z_6^{(2)})^{-1/2}
~.~
\eea
Accordingly, we can summarize that the birational transformation in \eref{es10a11}
identifies the Casimirs and the Hamiltonians as well as the spectral curve and the Poisson structure formed by the cluster variables of the two dimer integrable systems.
We thus identify the two integrable systems
corresponding to Model 2 and 4a to be birationally equivalent 
under \eref{es10a11}.

In general, we conjecture that whenever two toric Calabi-Yau 3-folds and their corresponding toric varieties are related by a birational transformation, 
there exists a pair of associated dimer models
whose respective dimer integrable systems are birationally equivalent.
Here, we define birational equivalence as an identification of the Casimirs and the Hamiltonians as well as the spectral curve and the Poisson structure formed by the cluster variables of the two dimer integrable systems.
We anticipate that this observation extends also to the corresponding quantum integrable systems.
Moreover, we can report here that this observation has been confirmed
for dimer models corresponding to more general toric Calabi-Yau 3-folds including those with toric diagrams that are not reflexive polygons. 
We also note here that the birational transformations identifying zig-zag loops and face loops between two different dimer models
have interesting interpretations as deformations of the associated $4d$ $\mathcal{N}=1$ gauge theories \cite{Ghim:2025zhs, Ghim:2024asj, Franco:2023flw, Arias-Tamargo:2024fjt, higashitani2022deformations}.
We also identify the birational transformations of the toric diagrams
to be related to Hanany-Witten moves in the dual $(p,q)$-web diagrams associated to $5d$ $\mathcal{N}=1$ gauge theories \cite{Franco:2023flw,Franco:2023mkw, Lee:2024bqg,Arias-Tamargo:2024fjt,CarrenoBolla:2024fxy}.
This connection also paves the way to further understand generalized toric polytopes (GTP) \cite{Franco:2023flw,Franco:2023mkw, Lee:2024bqg,Arias-Tamargo:2024fjt,CarrenoBolla:2024fxy} that appear in the study of $5d$ $\mathcal{N}=1$ gauge theories.
We plan to report further on these findings in the near future.

The authors would like to thank J. Bao, S. Franco, D. Ghim, Y.-H. He, S. Jeong, K. Lee and M. Yamazaki for discussions.
R.-K. S. is supported by an Outstanding Young Scientist Grant (RS-2025-00516583) of the National Research Foundation of Korea (NRF).
He is also partly supported by the BK21 Program (``Next Generation Education Program for Mathematical Sciences'', 4299990414089) funded by the Ministry of Education in Korea and the National Research Foundation of Korea (NRF).
N. L. is supported by the Institute of Basic Science (IBS-R003-D1).


\bibliographystyle{jhep}
\bibliography{mybib}

\providecommand{\href}[2]{#2}\begingroup\raggedright\begin{thebibliography}{10}

\bibitem{mori1}
S.~Mori, \emph{Projective manifolds with ample tangent bundles}, {\emph{Annals
  of Mathematics} {\bf 110} (1979) 593--606}.

\bibitem{mori2}
S.~Mori, \emph{Threefolds whose canonical bundles are not numerically
  effective}, {\emph{Annals of Mathematics} {\bf 116} (1982) 133--176}.

\bibitem{mori3}
S.~Mori, \emph{Flip theorem and the existence of minimal models for 3-folds},
  {\emph{Journal of the American Mathematical Society} {\bf 1} (1988)
  117--253}.

\bibitem{Kawamata}
Y.~Kawamata, \emph{Pluricanonical systems on minimal algebraic varieties},
  \href{http://dx.doi.org/10.1007/BF01388524}{\emph{Inventiones mathematicae}
  {\bf 79} (1985) 567--588}.

\bibitem{zbMATH00232903}
J.~Koll{\'a}r, Y.~Miyaoka and S.~Mori, \emph{Rationally connected varieties},
  {\emph{J. Algebr. Geom.} {\bf 1} (1992) 429--448}.

\bibitem{Kollar_Mori_1998}
J.~Kollár and S.~Mori, \emph{Birational Geometry of Algebraic Varieties}.
\newblock Cambridge Tracts in Mathematics. Cambridge University Press, 1998.

\bibitem{birkar2010existence}
C.~Birkar, P.~Cascini, C.~Hacon and J.~McKernan, \emph{Existence of minimal
  models for varieties of log general type}, {\emph{Journal of the American
  Mathematical Society} {\bf 23} (2010) 405--468}.

\bibitem{akhtar2012minkowski}
M.~Akhtar, T.~Coates, S.~Galkin, A.~M. Kasprzyk et~al., \emph{Minkowski
  polynomials and mutations}, {\emph{SIGMA. Symmetry, Integrability and
  Geometry: Methods and Applications} {\bf 8} (2012) 094}.

\bibitem{gross2013birational}
M.~Gross, P.~Hacking and S.~Keel, \emph{Birational geometry of cluster
  algebras},  \href{http://arxiv.org/abs/1309.2573}{{\tt 1309.2573}}.

\bibitem{coates2016quantum}
T.~Coates, A.~Corti, S.~Galkin and A.~Kasprzyk, \emph{Quantum periods for
  3--dimensional fano manifolds}, {\emph{Geometry \& Topology} {\bf 20} (2016)
  103--256}.

\bibitem{kasprzyk2017minimality}
A.~Kasprzyk, B.~Nill and T.~Prince, \emph{Minimality and mutation-equivalence
  of polygons},  in \emph{Forum of mathematics, Sigma}, vol.~5, p.~e18,
  Cambridge University Press, 2017.

\bibitem{coates2021maximally}
T.~Coates, A.~M. Kasprzyk, G.~Pitton and K.~Tveiten, \emph{Maximally mutable
  laurent polynomials}, {\emph{Proceedings of the Royal Society A} {\bf 477}
  (2021) 20210584}.

\bibitem{Coates2022MirrorSL}
T.~Coates, L.~Heuberger and A.~M. Kasprzyk, \emph{Mirror symmetry, laurent
  inversion and the classification of $\mathbb{Q}$-fano threefolds},
  \href{http://arxiv.org/abs/2210.07328}{{\tt 2210.07328}}.

\bibitem{Ghim:2024asj}
D.~Ghim, M.~Kho and R.-K. Seong, \emph{{Combinatorial and algebraic mutations
  of toric Fano 3-folds and mass deformations of 2d(0,2) quiver gauge
  theories}}, \href{http://dx.doi.org/10.1103/PhysRevD.110.086001}{\emph{Phys.
  Rev. D} {\bf 110} (2024) 086001},
  [\href{http://arxiv.org/abs/2407.19924}{{\tt 2407.19924}}].

\bibitem{Ghim:2025zhs}
D.~Ghim, M.~Kho and R.-K. Seong, \emph{{Birational transformations and 2d (0,
  2) quiver gauge theories beyond toric Fano 3-folds}},
  \href{http://dx.doi.org/10.1007/JHEP06(2025)032}{\emph{JHEP} {\bf 06} (2025)
  032}, [\href{http://arxiv.org/abs/2502.08741}{{\tt 2502.08741}}].

\bibitem{KENYON1997591}
R.~Kenyon, \emph{Local statistics of lattice dimers},
  \href{http://dx.doi.org/https://doi.org/10.1016/S0246-0203(97)80106-9}{\emph{Annales
  de l'Institut Henri Poincare (B) Probability and Statistics} {\bf 33} (1997)
  591--618}.

\bibitem{kenyon2003introductiondimermodel}
R.~Kenyon, \emph{An introduction to the dimer model},  2003.

\bibitem{Hanany:2005ve}
A.~Hanany and K.~D. Kennaway, \emph{{Dimer models and toric diagrams}},
  \href{http://arxiv.org/abs/hep-th/0503149}{{\tt hep-th/0503149}}.

\bibitem{Franco:2005rj}
S.~Franco, A.~Hanany, K.~D. Kennaway, D.~Vegh and B.~Wecht, \emph{{Brane dimers
  and quiver gauge theories}},
  \href{http://dx.doi.org/10.1088/1126-6708/2006/01/096}{\emph{JHEP} {\bf 01}
  (2006) 096}, [\href{http://arxiv.org/abs/hep-th/0504110}{{\tt
  hep-th/0504110}}].

\bibitem{Franco:2005sm}
S.~Franco, A.~Hanany, D.~Martelli, J.~Sparks, D.~Vegh and B.~Wecht,
  \emph{{Gauge theories from toric geometry and brane tilings}},
  \href{http://dx.doi.org/10.1088/1126-6708/2006/01/128}{\emph{JHEP} {\bf 01}
  (2006) 128}, [\href{http://arxiv.org/abs/hep-th/0505211}{{\tt
  hep-th/0505211}}].

\bibitem{Klebanov:1998hh}
I.~R. Klebanov and E.~Witten, \emph{{Superconformal field theory on
  three-branes at a Calabi-Yau singularity}},
  \href{http://dx.doi.org/10.1016/S0550-3213(98)00654-3}{\emph{Nucl. Phys. B}
  {\bf 536} (1998) 199--218}, [\href{http://arxiv.org/abs/hep-th/9807080}{{\tt
  hep-th/9807080}}].

\bibitem{Gubser:1998ia}
S.~Gubser, N.~Nekrasov and S.~Shatashvili, \emph{{Generalized conifolds and
  4-Dimensional N=1 SuperConformal Field Theory}},
  \href{http://dx.doi.org/10.1088/1126-6708/1999/05/003}{\emph{JHEP} {\bf 05}
  (1999) 003}, [\href{http://arxiv.org/abs/hep-th/9811230}{{\tt
  hep-th/9811230}}].

\bibitem{Bianchi:2014qma}
M.~Bianchi, S.~Cremonesi, A.~Hanany, J.~F. Morales, D.~Ricci~Pacifici and R.-K.
  Seong, \emph{{Mass-deformed Brane Tilings}},
  \href{http://dx.doi.org/10.1007/JHEP10(2014)027}{\emph{JHEP} {\bf 10} (2014)
  027}, [\href{http://arxiv.org/abs/1408.1957}{{\tt 1408.1957}}].

\bibitem{Franco:2023tyf}
S.~Franco, D.~Ghim, G.~P. Goulas and R.-K. Seong, \emph{{Mass deformations of
  brane brick models}},
  \href{http://dx.doi.org/10.1007/JHEP09(2023)176}{\emph{JHEP} {\bf 09} (2023)
  176}, [\href{http://arxiv.org/abs/2307.03220}{{\tt 2307.03220}}].

\bibitem{higashitani2022deformations}
A.~Higashitani, Y.~Nakajima et~al., \emph{Deformations of dimer models},
  {\emph{SIGMA. Symmetry, Integrability and Geometry: Methods and Applications}
  {\bf 18} (2022) 030}.

\bibitem{fulton1993introduction}
W.~Fulton, \emph{Introduction to toric varieties}.
\newblock No.~131. Princeton university press, 1993.

\bibitem{Leung:1997tw}
N.~C. Leung and C.~Vafa, \emph{{Branes and toric geometry}},
  \href{http://dx.doi.org/10.4310/ATMP.1998.v2.n1.a4}{\emph{Adv. Theor. Math.
  Phys.} {\bf 2} (1998) 91--118},
  [\href{http://arxiv.org/abs/hep-th/9711013}{{\tt hep-th/9711013}}].

\bibitem{kasteleyn1967graph}
P.~Kasteleyn, \emph{Graph theory and crystal physics}, {\emph{Graph theory and
  theoretical physics} (1967) 43--110}.

\bibitem{goncharov2012dimersclusterintegrablesystems}
A.~B. Goncharov and R.~Kenyon, \emph{Dimers and cluster integrable systems},
  2012.

\bibitem{EagerFrancoSchaeffer}
R.~Eager, S.~Franco and K.~Schaeffer, \emph{Dimer models and integrable
  systems}, \href{http://dx.doi.org/10.1007/JHEP06(2012)106}{\emph{Journal of
  High Energy Physics} {\bf 2012} (2012) 106}.

\bibitem{Bershtein:2017swf}
M.~Bershtein, P.~Gavrylenko and A.~Marshakov, \emph{{Cluster integrable
  systems, $q$-Painlev\'e equations and their quantization}},
  \href{http://dx.doi.org/10.1007/JHEP02(2018)077}{\emph{JHEP} {\bf 02} (2018)
  077}, [\href{http://arxiv.org/abs/1711.02063}{{\tt 1711.02063}}].

\bibitem{Marshakov:2019vnz}
A.~Marshakov and M.~Semenyakin, \emph{{Cluster integrable systems and spin
  chains}}, \href{http://dx.doi.org/10.1007/JHEP10(2019)100}{\emph{JHEP} {\bf
  10} (2019) 100}, [\href{http://arxiv.org/abs/1905.09921}{{\tt 1905.09921}}].

\bibitem{Huang:2020neq}
M.-x. Huang, Y.~Sugimoto and X.~Wang, \emph{{Quantum periods and spectra in
  dimer models and Calabi-Yau geometries}},
  \href{http://dx.doi.org/10.1007/JHEP09(2020)168}{\emph{JHEP} {\bf 09} (2020)
  168}, [\href{http://arxiv.org/abs/2006.13482}{{\tt 2006.13482}}].

\bibitem{Lee:2023wbf}
N.~Lee, \emph{{New dimer integrable systems and defects in five dimensional
  gauge theory}}, \href{http://dx.doi.org/10.1007/JHEP12(2024)050}{\emph{JHEP}
  {\bf 12} (2024) 050}, [\href{http://arxiv.org/abs/2312.13133}{{\tt
  2312.13133}}].

\bibitem{Lee:2024bqg}
K.~Lee and N.~Lee, \emph{{Dimers for type D relativistic Toda model}},
  \href{http://dx.doi.org/10.1007/JHEP09(2024)198}{\emph{JHEP} {\bf 09} (2024)
  198}, [\href{http://arxiv.org/abs/2406.00925}{{\tt 2406.00925}}].

\bibitem{Douglas:1997de}
M.~R. Douglas, B.~R. Greene and D.~R. Morrison, \emph{{Orbifold resolution by
  D-branes}},
  \href{http://dx.doi.org/10.1016/S0550-3213(97)00517-8}{\emph{Nucl. Phys. B}
  {\bf 506} (1997) 84--106}, [\href{http://arxiv.org/abs/hep-th/9704151}{{\tt
  hep-th/9704151}}].

\bibitem{Douglas:1996sw}
M.~R. Douglas and G.~W. Moore, \emph{{D-branes, quivers, and ALE instantons}},
  \href{http://arxiv.org/abs/hep-th/9603167}{{\tt hep-th/9603167}}.

\bibitem{Feng:2000mi}
B.~Feng, A.~Hanany and Y.-H. He, \emph{{D-brane gauge theories from toric
  singularities and toric duality}},
  \href{http://dx.doi.org/10.1016/S0550-3213(00)00699-4}{\emph{Nucl. Phys. B}
  {\bf 595} (2001) 165--200}, [\href{http://arxiv.org/abs/hep-th/0003085}{{\tt
  hep-th/0003085}}].

\bibitem{Feng:2001xr}
B.~Feng, A.~Hanany and Y.-H. He, \emph{{Phase structure of D-brane gauge
  theories and toric duality}},
  \href{http://dx.doi.org/10.1088/1126-6708/2001/08/040}{\emph{JHEP} {\bf 08}
  (2001) 040}, [\href{http://arxiv.org/abs/hep-th/0104259}{{\tt
  hep-th/0104259}}].

\bibitem{Hori:2000ck}
K.~Hori, A.~Iqbal and C.~Vafa, \emph{{D-branes and mirror symmetry}},
  \href{http://arxiv.org/abs/hep-th/0005247}{{\tt hep-th/0005247}}.

\bibitem{Hori:2000kt}
K.~Hori and C.~Vafa, \emph{{Mirror symmetry}},
  \href{http://arxiv.org/abs/hep-th/0002222}{{\tt hep-th/0002222}}.

\bibitem{Feng:2005gw}
B.~Feng, Y.-H. He, K.~D. Kennaway and C.~Vafa, \emph{{Dimer models from mirror
  symmetry and quivering amoebae}},
  \href{http://dx.doi.org/10.4310/ATMP.2008.v12.n3.a2}{\emph{Adv. Theor. Math.
  Phys.} {\bf 12} (2008) 489--545},
  [\href{http://arxiv.org/abs/hep-th/0511287}{{\tt hep-th/0511287}}].

\bibitem{Davey:2010px}
J.~Davey, A.~Hanany and R.-K. Seong, \emph{{Counting Orbifolds}},
  \href{http://dx.doi.org/10.1007/JHEP06(2010)010}{\emph{JHEP} {\bf 06} (2010)
  010}, [\href{http://arxiv.org/abs/1002.3609}{{\tt 1002.3609}}].

\bibitem{Hanany:2010ne}
A.~Hanany and R.-K. Seong, \emph{{Symmetries of Abelian Orbifolds}},
  \href{http://dx.doi.org/10.1007/JHEP01(2011)027}{\emph{JHEP} {\bf 01} (2011)
  027}, [\href{http://arxiv.org/abs/1009.3017}{{\tt 1009.3017}}].

\bibitem{Hanany:2012hi}
A.~Hanany and R.-K. Seong, \emph{{Brane Tilings and Reflexive Polygons}},
  \href{http://dx.doi.org/10.1002/prop.201200008}{\emph{Fortsch. Phys.} {\bf
  60} (2012) 695--803}, [\href{http://arxiv.org/abs/1201.2614}{{\tt
  1201.2614}}].

\bibitem{Feng:2002fv}
B.~Feng, S.~Franco, A.~Hanany and Y.-H. He, \emph{{UnHiggsing the del Pezzo}},
  \href{http://dx.doi.org/10.1088/1126-6708/2003/08/058}{\emph{JHEP} {\bf 08}
  (2003) 058}, [\href{http://arxiv.org/abs/hep-th/0209228}{{\tt
  hep-th/0209228}}].

\bibitem{Witten:1993yc}
E.~Witten, \emph{{Phases of N=2 theories in two-dimensions}},
  \href{http://dx.doi.org/10.1016/0550-3213(93)90033-L}{\emph{Nucl. Phys. B}
  {\bf 403} (1993) 159--222}, [\href{http://arxiv.org/abs/hep-th/9301042}{{\tt
  hep-th/9301042}}].

\bibitem{Jejjala:2010vb}
V.~Jejjala, S.~Ramgoolam and D.~Rodriguez-Gomez, \emph{{Toric CFTs, Permutation
  Triples and Belyi Pairs}},
  \href{http://dx.doi.org/10.1007/JHEP03(2011)065}{\emph{JHEP} {\bf 03} (2011)
  065}, [\href{http://arxiv.org/abs/1012.2351}{{\tt 1012.2351}}].

\bibitem{Hanany:2015tgh}
A.~Hanany, V.~Jejjala, S.~Ramgoolam and R.-K. Seong, \emph{{Consistency and
  Derangements in Brane Tilings}},
  \href{http://dx.doi.org/10.1088/1751-8113/49/35/355401}{\emph{J. Phys. A}
  {\bf 49} (2016) 355401}, [\href{http://arxiv.org/abs/1512.09013}{{\tt
  1512.09013}}].

\bibitem{Franco:2023flw}
S.~Franco and R.-K. Seong, \emph{{Twin theories, polytope mutations and quivers
  for GTPs}}, \href{http://dx.doi.org/10.1007/JHEP07(2023)034}{\emph{JHEP} {\bf
  07} (2023) 034}, [\href{http://arxiv.org/abs/2302.10951}{{\tt 2302.10951}}].

\bibitem{Arias-Tamargo:2024fjt}
G.~Arias-Tamargo, S.~Franco and D.~Rodr\'\i{}guez-G\'omez, \emph{{The geometry
  of GTPs and 5d SCFTs}},
  \href{http://dx.doi.org/10.1007/JHEP07(2024)159}{\emph{JHEP} {\bf 07} (2024)
  159}, [\href{http://arxiv.org/abs/2403.09776}{{\tt 2403.09776}}].

\bibitem{Franco:2023mkw}
S.~Franco and D.~Rodriguez-Gomez, \emph{{Quiver tails and brane webs}},
  \href{http://dx.doi.org/10.1007/JHEP10(2024)118}{\emph{JHEP} {\bf 10} (2024)
  118}, [\href{http://arxiv.org/abs/2310.10724}{{\tt 2310.10724}}].

\bibitem{CarrenoBolla:2024fxy}
I.~Carre\~no Bolla, S.~Franco and D.~Rodr\'\i{}guez-G\'omez, \emph{{The 5d
  tangram: brane webs, 7-branes and primitive T-cones}},
  \href{http://dx.doi.org/10.1007/JHEP05(2025)175}{\emph{JHEP} {\bf 05} (2025)
  175}, [\href{http://arxiv.org/abs/2411.01510}{{\tt 2411.01510}}].

\end{thebibliography}\endgroup

\end{document}